\documentclass[aps,prl,showpacs,twocolumn,floatfix]{revtex4}
\usepackage{epsf}
\usepackage{amssymb,amsmath}
\usepackage[dvips]{color}
\usepackage{graphicx}

\newcommand{\acr}[1]{\hat{a} _{#1} ^{\dag}}
\newcommand{\aan}[1]{\hat{a} _{#1} }

\begin{document}

\title{Bose-Einstein Condensates in Rotating Lattices} \author{Rajiv
  Bhat and M.~J. Holland} \affiliation{JILA, National Institute of
  Standards and Technology and Department of Physics, University of
  Colorado Boulder, CO 80309} \author{L.~D. Carr} \affiliation{
  Physics Department, Colorado School of Mines, Golden, CO 80401}
  \date{\today}

\begin{abstract}
  Strongly interacting bosons in a two-dimensional rotating square lattice are
  investigated via a modified Bose-Hubbard Hamiltonian.  Such a system
  corresponds to a rotating lattice potential imprinted on a trapped
  Bose-Einstein condensate. Second-order quantum phase transitions
  between states of different symmetries are observed at discrete
  rotation rates. For the square lattice we study, there are four
  possible ground-state symmetries.
\end{abstract}

\pacs{}

\maketitle

\begin{figure}[t]
  \includegraphics[totalheight=5.0cm,width=7.8cm]{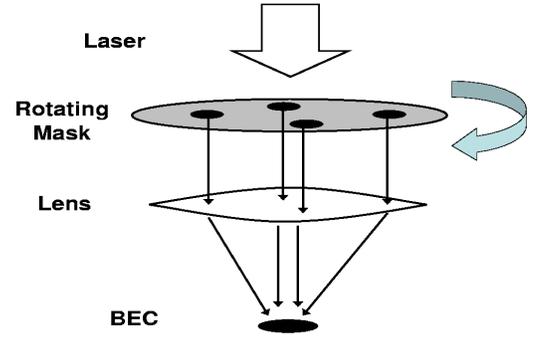}
  \caption{(color online) Sketch of a rotating lattice. A laser beam
    passes through a rotating mask. A lens focuses the resulting
    lattice beam profile onto a stationary, trapped BEC. The lattice
    potential is imprinted via off-resonant interactions between the
    laser and the atoms.\label{fig:1}}
\vspace{-0.03\textwidth}
\end{figure}
One of the most exciting recent advances in the field of ultracold
quantum gases has been the realization of quantum phase transitions in
Bose-Einstein condensates (BEC's) trapped in a
lattice~\cite{greiner2002}, in particular the superfluid to Mott
insulator transition~\cite{fisher1989}. Quantum phase transitions
typically occur when a topological change happens in the ground state
symmetry as a function of a Hamiltonian parameter at zero
temperature~\cite{sachdev1999}. They play an important role in many
models, including the quantum Ising model, quantum rotors, and the
Bose-Hubbard model~\cite{sachdev1999}. The latter very accurately
describes experimental observations of the superfluid-Mott transition
in BEC's~\cite{jaksch1998,greiner2002b}.

A second fruitful development in BEC's has been the study of rotating
systems in harmonic traps, in which an Abrikosov lattice of quantized
vortices appears for sufficient rotation~\cite{madison2000}. Rotating
BEC's have also been predicted to realize an analog of the fractional
quantum Hall effect~\cite{wilkin2000}.  More generally, the density,
temperature, lattice structure, strength and symmetry of
interactions in ultracold quantum gases are precisely and dynamically
controllable in experiments, with no impurities or disorder.  Thus
these systems are accurately described by relatively simple model
Hamiltonians. In this sense they are a playground for quantum many
body theory.

In this Letter, we pull together these two cutting edge areas of study
in BEC's, i.e., rotating systems and quantum phase transitions in
lattices.  Specifically, we derive and solve a modified Bose-Hubbard
Hamiltonian for rotating BEC's trapped in a two-dimensional(2D) square
lattice potential with variable filling factor. In addition to the
superfluid-Mott transition, which can be driven by hopping and/or
rotation, we find that a quantum phase transition occurs each time the
rotational symmetry of the ground state changes. The possible discrete
values of the rotational symmetry are determined by the geometry of
the underlying lattice. Fig.~\ref{fig:1} is a sketch of one way to
make a 2D rotating lattice, as has already been experimentally
realized at JILA~\cite{schweikhard2005}. Some theoretical aspects of
this problem for large filling factors, such as vortex
pinning~\cite{reijnders2005}, structural phase transitions of vortex
matter~\cite{pu2005}, and the single vortex problem~\cite{wuC2004}
have been studied previously. In contrast, we will consider filling
factors of less than unity.  In particular, we treat small, finite
systems where exact solution via diagonalization of the Hamiltonian in
a truncated Hilbert space is insightful. Recall that, in most
experiments, condensates typically are comprised of $~10^6$ atoms, 3D
optical lattices consist of $~100^3$ sites spaced at 300--500~nm
~\cite{greiner2002}, and exact solutions for systems of this size are
not tractable numerically. Our approach builds up a scalable picture
of the lattice physics from the microscopic interaction and lattice
structure.

Consider bosons interacting via an effective 2D two-body contact potential of
strength $g$ in a 2D lattice potential
$V^{\mathrm{lat}}(x,y)=V^{\mathrm{lat}}(x+j d,y + k d)$ which rotates
about the $z$~axis, where $j,k$ are integers and $d$ is the lattice
constant. In the rotating frame, the Hamiltonian is
\begin{align}
  \hat{H}= &\int\!\!d^2r\, \hat{\Phi}^{\dag}
  \left[-\frac{\hbar^2}{2M}\nabla^2
    +\frac{g}{2}\hat{\Phi}^{\dag}\hat{\Phi} + V^{\mathrm{lat}}-\Omega
    L_z\right] \hat{\Phi}\,, \label{eqn:h}\\
  L_z \equiv & -i\hbar( x\partial_y - y
\partial_x ), \label{eqn:lz}
\end{align}
with rotation frequency $\Omega$ and atomic mass $M$. The field
operator, $\hat{\Phi}(x,y)$, obeys the usual bosonic commutation
relations. Using a Wannier basis, $W _i (x,y)$, the field operator can
be expanded in terms of bosonic operators $\hat{a}_i$,
\begin{equation}
  \hat{\Phi}(x,y) = \textstyle \sum _i
  \hat{a}_{i} W _i (x,y)\,.
\end{equation}
The usual single-band Bose-Hubbard model~\cite{fisher1989,sachdev1999}
is obtained via the tight binding and lowest band
approximations~\cite{rey2003}. The rotational part $\hat{H}_L\equiv
\int d^2r\, \hat{\Phi}^{\dagger}\Omega L_z \hat{\Phi}$ of
Eq.~\eqref{eqn:h} becomes
\begin{align}
  \hat{H}_L &= -i\hbar \Omega \textstyle\int \! d^2r \textstyle\sum
  _{<i,j>} \nonumber\\& \left[ \acr{j} \aan{i} (W _j ^{*}(x,y) (x
    \partial_y - y \partial_x)
    W _i (x,y))+ h.c. \right] \nonumber \\
  &\equiv i\hbar \Omega \textstyle\sum _{<i,j>} K
  _{ij}(\aan{i}\acr{j}-\acr{i}\aan{j})\,,
\end{align}
where $<\! i,j\!>$ indicates a sum over nearest neighbors. The general
form of $K _{ij}$ is
\begin{equation}
  K _{ij}= \beta(r _i r _j/d^2)\sin\alpha _{ij} \label{eqn:K}\,,
\end{equation}
where $r _i$ denotes the distance from the axis of rotation to the $i
^{th}$ site, $\alpha _{ij}$ is the angle subtended by the $i
^{\mathrm{th}}$ and the $j ^{\mathrm{th}}$ sites with respect to the
axis of rotation, and $\beta$ is a dimensionless constant
characterizing the lattice geometry and depth.

The rotating Bose-Hubbard Hamiltonian in the canonical ensemble is
\begin{align}
  \hat{H} = &-t \textstyle\sum _{<i,j>} (\aan{i}\acr{j}+\acr{i}\aan{j}
  ) \nonumber \\ & - i \hbar \Omega \textstyle\sum _{<i,j>}K _{ij}
  (\aan{i}\acr{j}-\acr{i}\aan{j} ) \nonumber \\
  & + \textstyle\frac{1}{2}U \textstyle\sum _{i} \hat{n} _{i} (\hat{n}
  _{i} -1)\,, \label{eqn:ham}
\end{align}
where $\hat{n} _{i}=\acr{i}\aan{i}$. The hopping energy $t$ and on-site
interaction energy $U$ have been explicitly calculated from $g$, $M$, etc.
elsewhere~\cite{rey2003}.  Equation~\eqref{eqn:ham} is the usual
Bose-Hubbard Hamiltonian with the addition of the second term, which
describes the rotation.

The expectation value of the current is an important ground state
observable for rotating quantum systems. In the Heisenberg picture the
current is given by
\begin{eqnarray}
  \langle \hat{J} _{ij} \rangle &=& (i/\hbar d)
  \langle [ \hat{n}_i, \hat{H}_{ij}]
  \rangle \label{eqn:current}\\
  &=& \frac{it}{\hbar d}
  \langle\aan{i} \acr{j}-\acr{i} \aan{j}\rangle
  -\frac{\Omega K _{ij}}{d}
  \langle\aan{i} \acr{j}+\acr{i}
  \aan{j}\rangle
  \,,\nonumber
\end{eqnarray}
where $\hat{H}_{ij}$ is the Hamiltonian for sites $i,j$ alone. In
Eq.~\eqref{eqn:current}, the first term is due to hopping while the
second is due to rotation. The current is conserved at each site $i$
in the rotating frame since the sum of $\langle \hat{J}_{ij} \rangle$
over all nearest neighbors $j$ is zero. Another useful observable is
the total current on the lattice boundary $ C $,
\begin{equation} 
  \Lambda \equiv (\hbar d/E_r )\textstyle\sum_{\langle
    i,j\rangle\in C }\langle \hat{J}_{ij} \rangle\,,\label{lambda}
\end{equation}
where we have scaled away the units via a ``recoil'' energy $E_r\equiv
\hbar^2/M d^2$ and all sums over $ C $ are taken with the same sign
convention as the helicity of $\Omega$.  We also define two
number-related observables: $n=\sum_i\langle \hat{n_i} \rangle$, the
average total number of atoms in the system, and
$\nu\equiv\textstyle\sum_i(\langle \hat{n_i}^2\rangle -\langle
\hat{n_i}\rangle^2)/\textstyle\sum_i\langle \hat{n}_i\rangle$, the
normalized variance.  Recall that $\nu=1$ for a coherent state, $\nu >
1$ for a phase-squeezed state, $\nu <1 $ for a number-squeezed state,
and $\nu=0$ for a single Fock state.
\begin{figure}[t]
\begin{center}
   \includegraphics[totalheight=5.5cm,width=8.3cm]{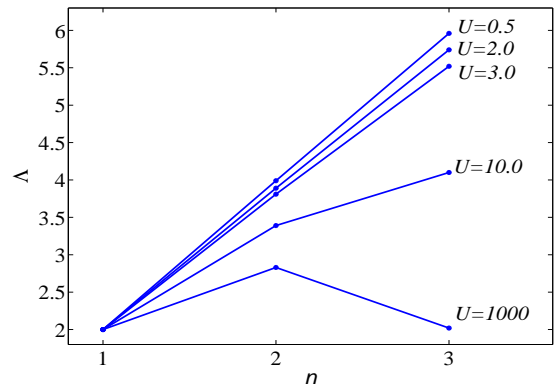}
    \vspace{-0.01\textwidth}
   \caption{(color online) The total dimensionless current $\Lambda$
     around a lattice unit cell is shown as a function of the average
     total number of atoms $n$ in the system for $\hbar \Omega K >t$.
     The curves from top to bottom are for increasing on-site
     interaction strength $U$ in units of the lattice recoil energy
     $E_r$.  For weak interactions, the atoms behave independently;
     for strong interactions, there is a particle-hole symmetry, as
     evident in the bottommost curve, where $n=1$ and $n=3$ have the
     same $\Lambda$. The connecting lines here are a guide to the
     eye.\label{fig:2}}
\end{center}
\vspace{-0.05\textwidth}
\end{figure}

The ground state wave function must be consistent with the discrete
rotational symmetry of the lattice.  This follows in general from the
point symmetry group $C_j$ of the lattice, which describes the
rotation of a $j$-fold symmetric object, e.g., a $j$-sided regular
polygon. The number of possible rotational symmetries is equal to $j$,
e.g., for the square lattice we consider $j=4$. The quantum mechanical
rotation operator about an angle $2\pi/j$, ${\cal R}(2\pi/j)$, then
commutes with the lattice potential part of the Hamiltonian. This
means that all non-degenerate ground-states are simultaneous
eigenstates of energy and of the discrete rotational symmetry
operator. Furthermore, $[{\cal R}(2\pi/j)]^j$, rotates the system
through an angle $2\pi$ and therefore must be the identity operator for
 the wave function to be single-valued. The eigenvalues of ${\cal
  R}(2\pi/j)$ can thus only take the values $e^{i2\pi m/j}$ where $m$
is an integer and $m\in\{0,j-1\}$. In the case where the total
particle number is commensurate with $j$, the eigenvalues of $[{\cal
  R}(2\pi/j)]$ can be degenerate, leading to interesting effects which
will be discussed later.

For simplicity, we will focus our discussion on the strongly
interacting case, achieved experimentally either by a Feshbach
resonance or by turning up the lattice potential height.  In this case
one can make a \emph{two-state approximation}~\cite{carr2005b} which
prevents multiple occupancy of the same site. Such a Fock space
eliminates the interaction term $U$ in Eq.~\eqref{eqn:h}.  However,
there are effective strong interactions due to atoms being unable to
cross each other; i.e., one has a system of hard core bosons.

Having defined the observables, we first consider the case of a single
square unit cell, i.e., a $2 \times 2$-site lattice.  In order to
assess the two-state approximation, we allow Fock states with up to
three atoms per site: however, we keep the average number of particles
per site at unity or below. We exactly diagonalize the Hamiltonian and
find the ground state. Due to the competition between the hopping and
rotational energy terms in Eq.~\eqref{eqn:ham}, rotation affects the
system only when $\hbar\Omega K> t$.  For a total number of atoms
$n\in \{1,2,3\}$, $m=0$ for $\hbar\Omega K<t$ and $m=1 $ for
$\hbar\Omega K>t$.  As shown in Fig.~\ref{fig:2}, for non-interacting
atoms ($U=0$) this corresponds to $\Lambda=0$ and $\Lambda=2 n$,
respectively. As the interatomic repulsion is increased, $\Lambda$
decreases to non-integer values.  Although $m$ is quantized, $\Lambda$
is not.  However, for $t/U\ll 1$ and $\hbar\Omega/U\ll 1$, i.e., for
very strong interactions, or a very strong lattice and small
rotation, the allowed values of $\Lambda$ return to those given by the
two-state approximation. The two-state approximation (zero or one atom
per site) is adequate to study strongly interacting systems.
\begin{figure}[t]
\begin{center}
   \includegraphics[totalheight=7.0cm,width=7.8cm]{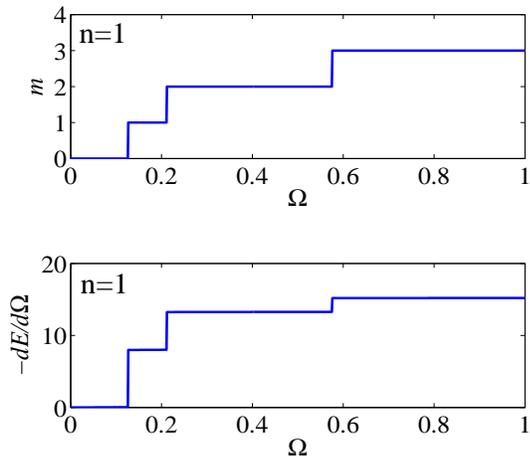}
    \vspace{-0.03\textwidth}
    \caption{ (color online) (a) index $m$ of the eigenvalue $e^{i2\pi
        m/4}$ of the discrete rotational symmetry operator $\cal R$ as
      a function of rotation rate, showing the discontinuous changes
      in the ground state symmetry. The rotation $\Omega$ is in units
      of $E_r/\hbar$, with total number of atoms $n=1$, hopping
      $t/E_r=1$, and $\beta=4.93\,t$ (as appropriate for a sinusoidal
      square lattice).  (b) The derivative of the ground state energy
      with respect to $\Omega$ in units of $\hbar$, showing discrete
      jumps at the points of level crossings.
     \label{fig:3}}
\end{center}
\vspace{-0.05\textwidth}
\end{figure}

Consider next a $4\times 4$-site square 2D lattice, which consists of
9 unit cells, in the two-state approximation. The numerical study
shows three main results. (1) Second order quantum phase transitions
occur each time the symmetry $m$ of the ground state changes.  (2)~As
the rotation rate increases, eventually the maximum current is
achieved corresponding to a phase difference of $\pi/2$ between the
wavefunction on adjacent sites along the perimeter. (3)~For higher
fillings, both the hopping $t$ and the rotation $\Omega$ can drive the
system through the extensively studied Mott-insulator/superfluid
transition ~\cite{fisher1989,jaksch1998,greiner2002,sachdev1999} (not
discussed further here).

\begin{figure}[t]
  \begin{minipage}[]{0.50\textwidth}
        \includegraphics[totalheight=3.8cm,width=4.2cm]{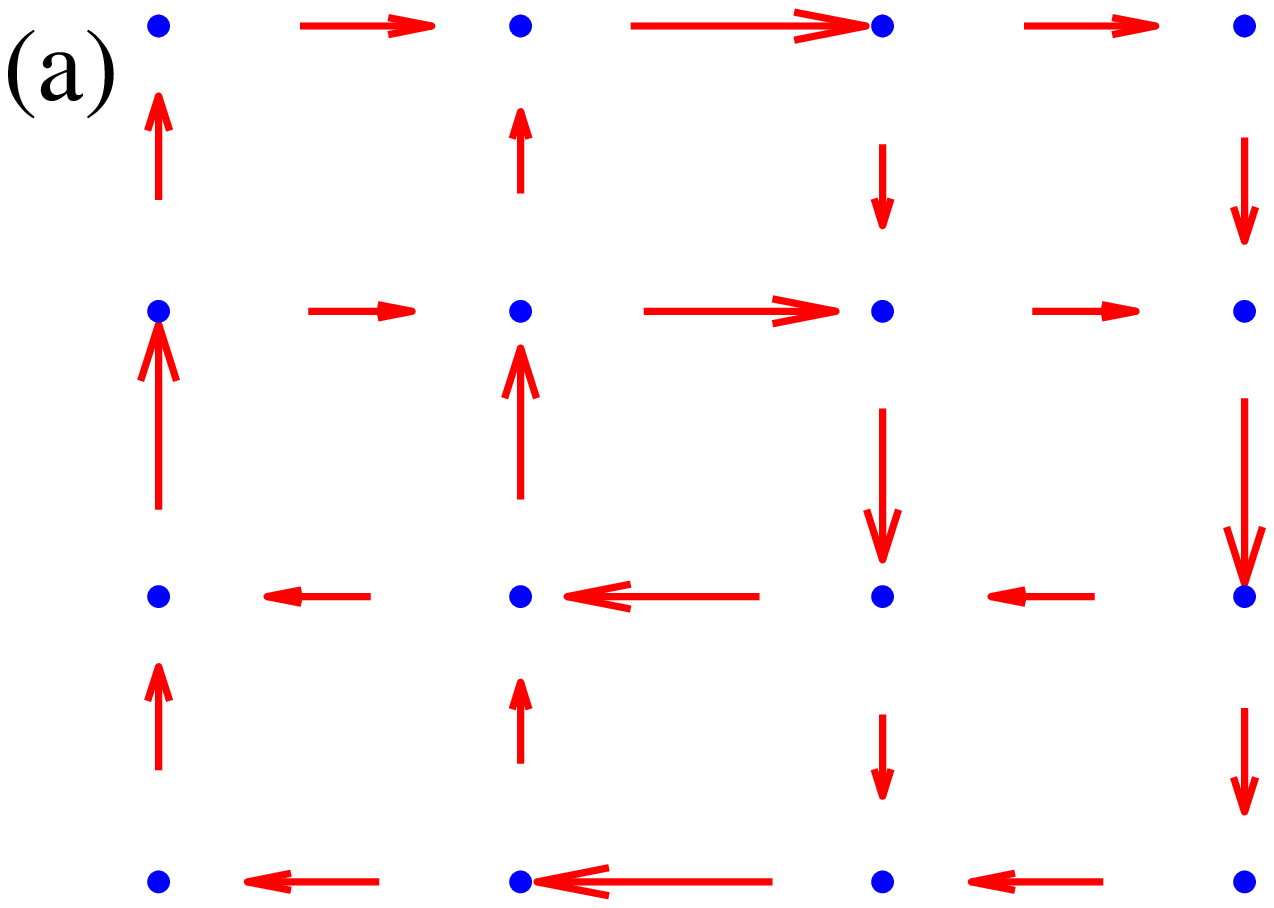}
        \hspace{-.75cm} 
        \includegraphics[totalheight=3.8cm,width=4.2cm]{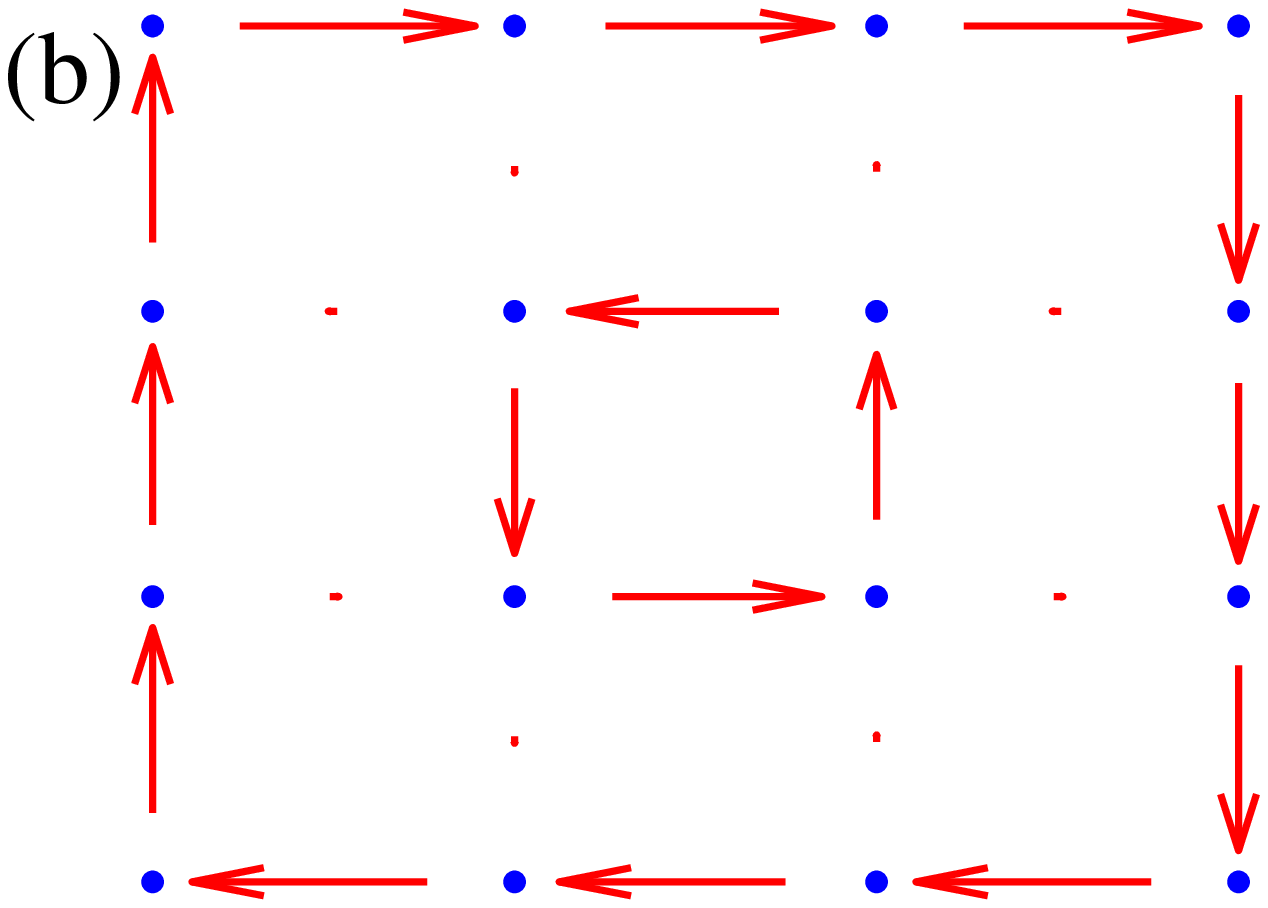}
    \vspace{-0.08\textwidth}
        \end{minipage}
    \begin{minipage}[]{0.50\textwidth}
       \includegraphics[totalheight=3.8cm,width=4.2cm]{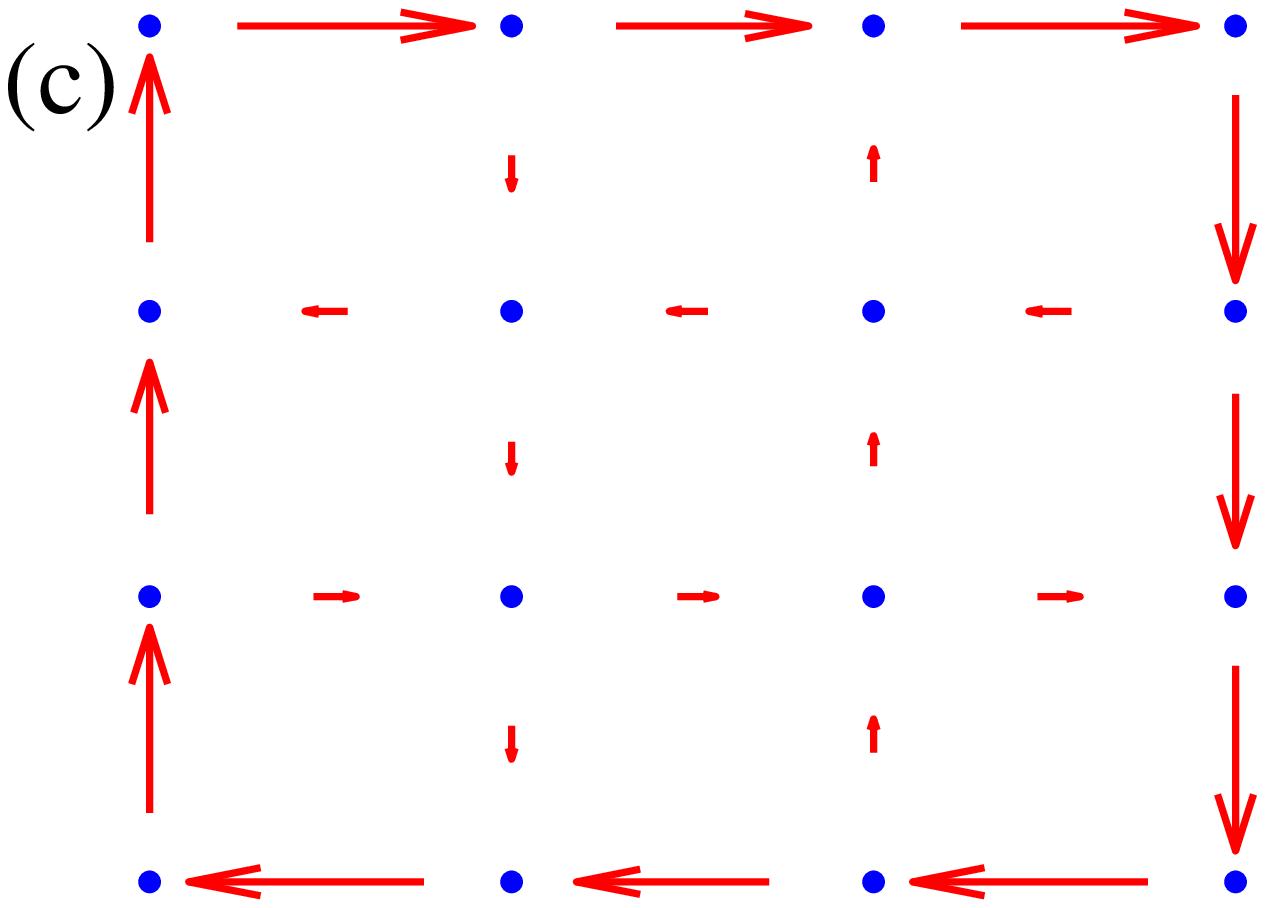}
       \hspace{-.75cm}
        \includegraphics[totalheight=3.8cm,width=4.2cm]{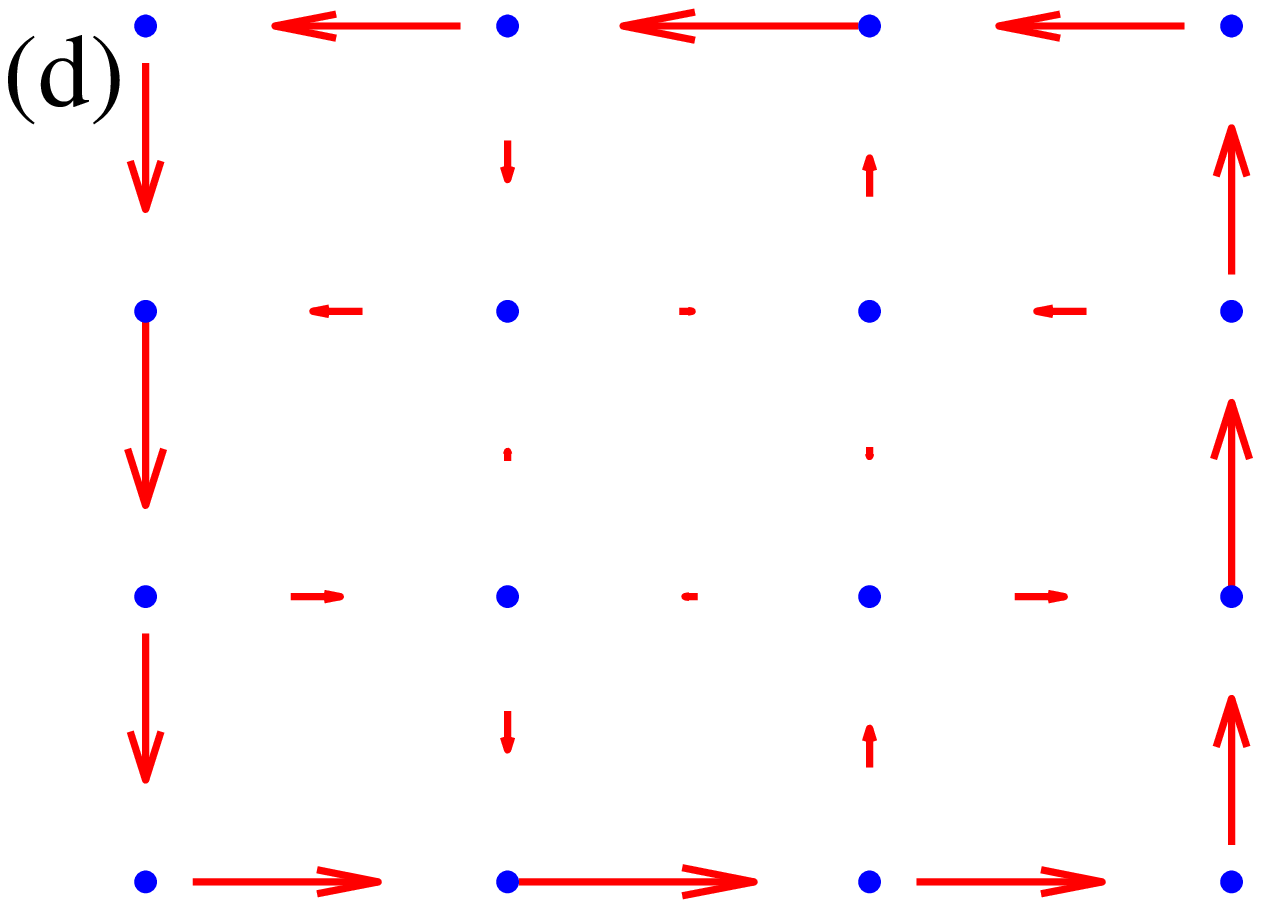}
      \vspace{-0.05\textwidth}
    \end{minipage}
    \caption{\label{fig:4} (color online) The red arrows indicate
      direction of current flow in the rotating frame for $t/E_r=1$
      and $n=1$ at increasing rotations on a $4\times 4$ lattice,
      corresponding to $m=0$, 1, 2, and~3. (a) $\hbar\Omega/E_r =
      0.1$; (b) $\hbar\Omega/E_r =0.2$; (c) $\hbar\Omega/E_r =0.4$;
      (d) $\hbar\Omega/E_r=0.8$.}
      \vspace{-0.035\textwidth}   
\end{figure}

First, consider the simplified case of one atom in the system. To
illustrate the QPTs, in Fig.~\ref{fig:3} is shown the eigenvalue index
$m$ and the derivative of the total energy $E\equiv \langle \hat{H}
\rangle $ with respect to $\Omega$. Exact energy level crossings are
observed as indicated by the discrete jumps in the energy derivative,
corresponding to each transition.  The corresponding circulation
patterns are shown in Fig.~\ref{fig:4}.  In Fig.~\ref{fig:4}(a), $m=0$
and rotation has not yet entered the system.  In the rotating frame,
the current seems to be flowing backwards, i.e., clockwise.  In
Fig.~\ref{fig:4}(b), $m=1$ and a single vortex enters the system and
rests at the center.  Fig.~\ref{fig:4}(c) shows a similar pattern
although $m=2$ in this case. Fig.~\ref{fig:4}(d) is analogous to a
tightly packed vortex lattice with $m=3$; the current directions in
the center with opposite sign of circulation to the flow around the
perimeter is a trivial result of packing four vortices together on a
2D square lattice with an antivortex at the center.

\begin{figure}[t]
  \centering
  \begin{minipage}[c]{0.50\textwidth}
    \includegraphics[totalheight=4.5cm,width=7.5cm]{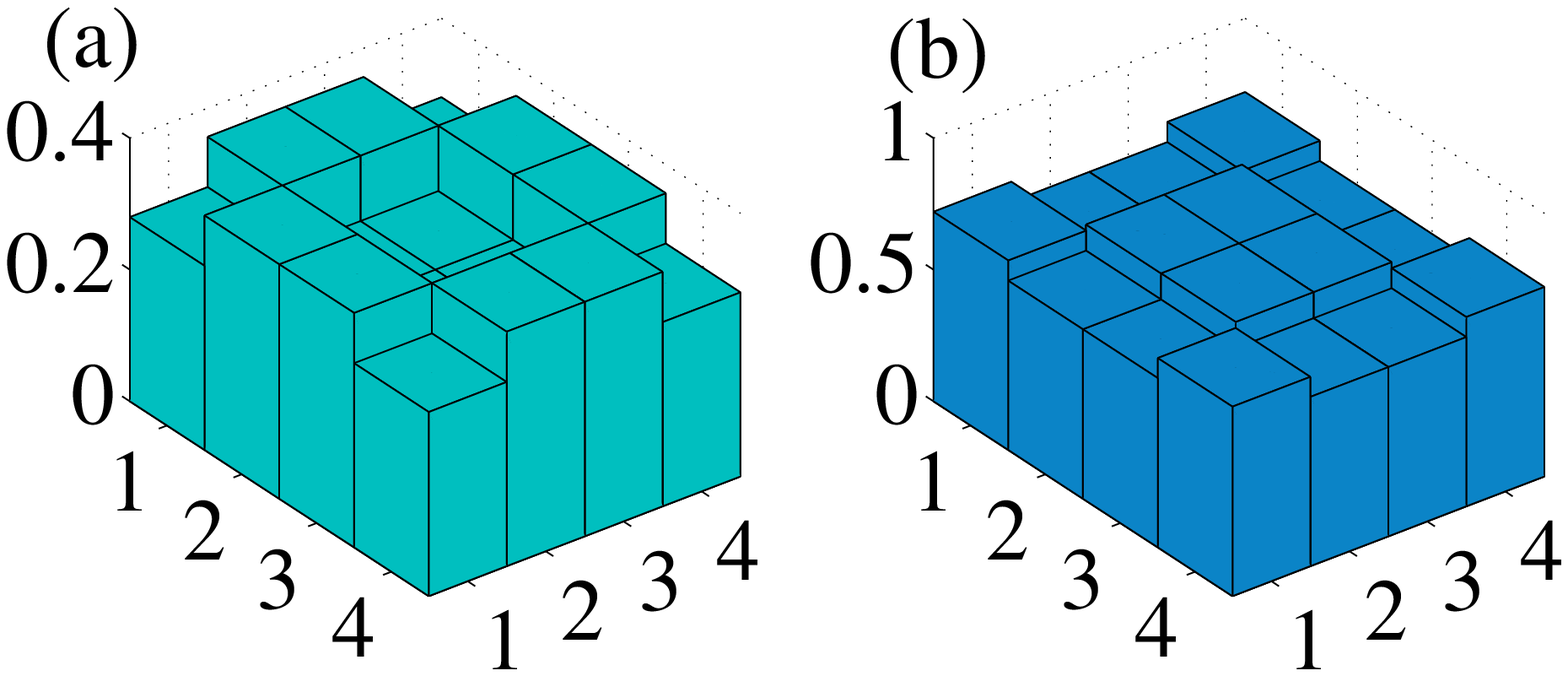}
    \vspace{-0.1\textwidth}
    \end{minipage}
    \begin{minipage}[c]{0.50\textwidth}
        \includegraphics[totalheight=3.8cm,width=8.5cm]{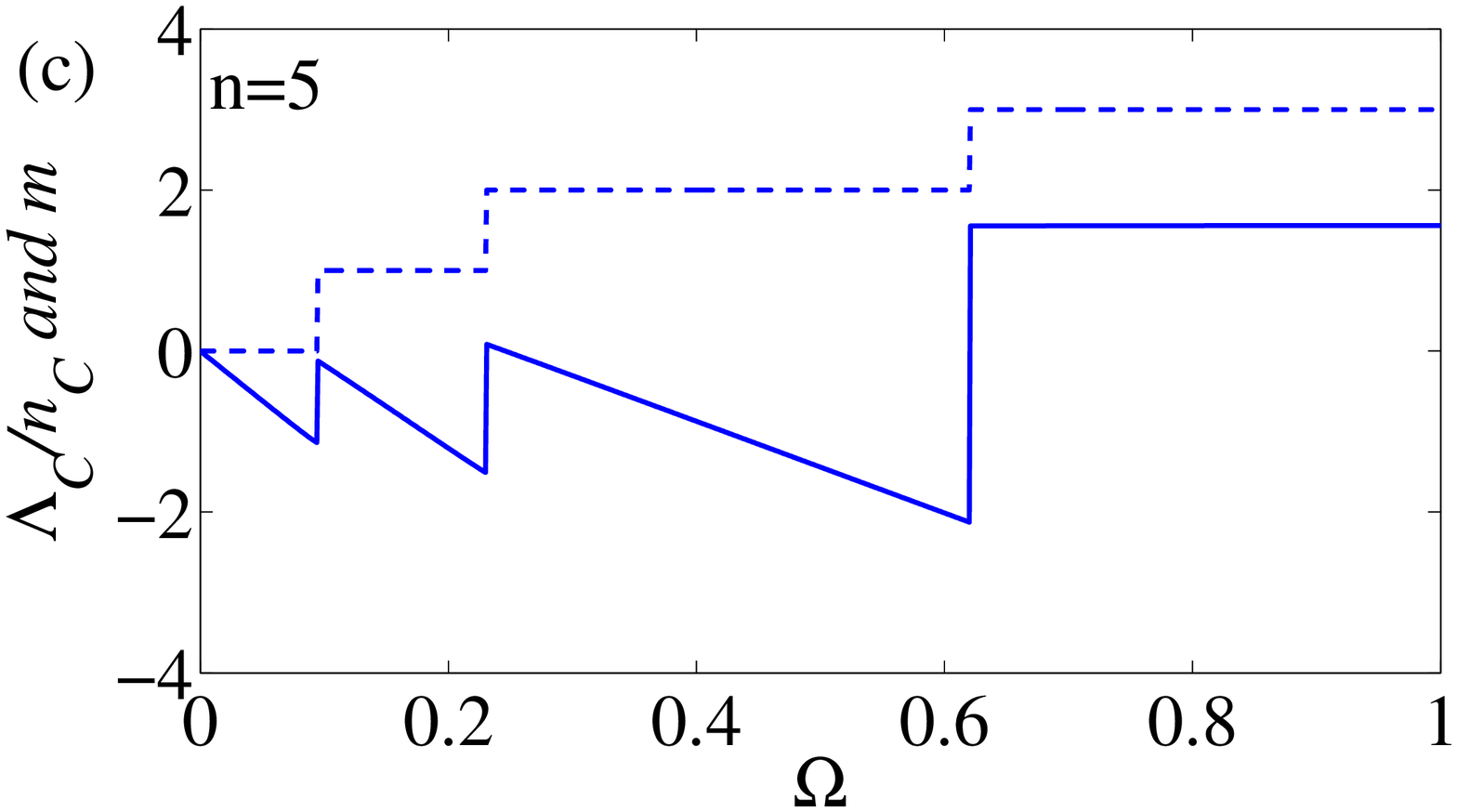}
         \vspace{-0.01\textwidth}
    \end{minipage}
    \begin{minipage}[c]{0.50\textwidth}
        \includegraphics[totalheight=3.8cm,width=8.5cm]{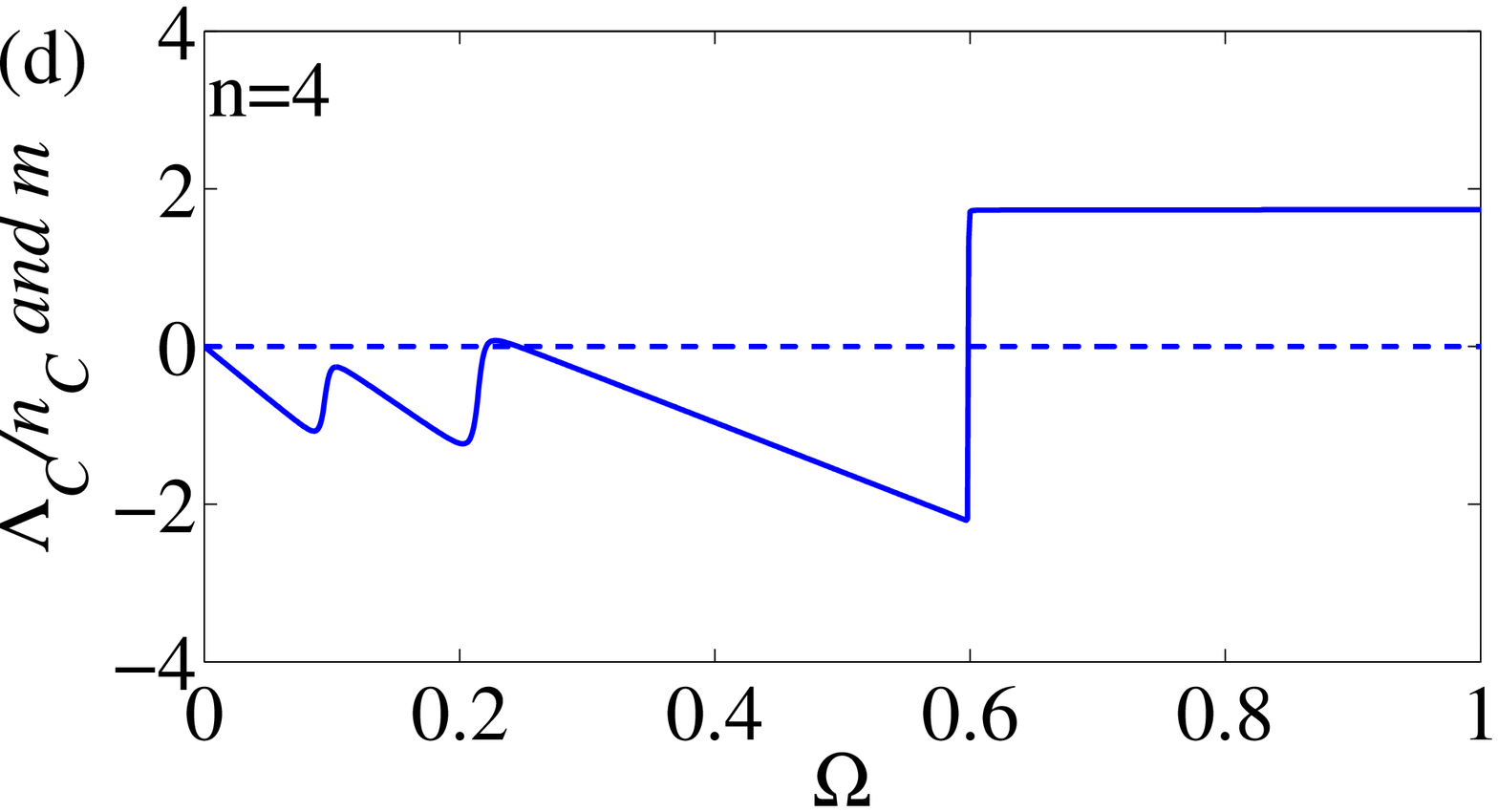}
    \end{minipage}
    \caption{ (color online) Observables for a $4 \times 4$ lattice
      with five particles and $t/E_r=1$.  (a) Site-dependent number
      density $n$ for $\hbar\Omega/E_r=0.1$.  There is noticeable
      depletion on the inner sites. (b) Site-dependent normalized
      variance $\nu$ for $\hbar\Omega/E_r =0.1$. Note that $\nu<1$, so
      this is a number-squeezed ground state. (c) The normalized,
      scaled current $\Lambda_C/n_C$ (solid curve) and $m$ (dashed
      curve), all on the boundary as a function of $\Omega$(in units
      of $E_r/\hbar$). Between each quantum phase transition, where
      the symmetry of the ground state changes abruptly as indicated
      by the jump in the value of $m$, the normalized current depends
      linearly on the rotation~$\Omega$. (d) is the same as (c) but
      describes four particles.
      \label{fig:5}} \vspace{-0.02\textwidth}
\end{figure}

Finally, these results can be extended for general filling. The main
difference from the single-atom case is the presence of effective
strong interactions due to the use of the two-state approximation. In
Fig.~\ref{fig:5} we illustrate the number density $n_d$, normalized
variance $\nu$, eigenvalue of discrete rotational symmetry $m$, and
the normalized current on the boundary $\Lambda/n_C$, all for a total
of five particles in the $4\times4$ square lattice. Note that $n_C$,
the number density along the perimeter, depends on $\Omega$.  In
Fig.~\ref{fig:5}(a)-(b) the site-dependence of $n_d$ and $\nu$ are
illustrated for $\hbar\Omega=t=E_r$.  In Fig.~\ref{fig:5}(c) one
observes that $\Lambda$ depends linearly on $\Omega$; this is
qualitatively a similar result to that of one atom.  However,
$\max(\Lambda_C/n_ C)\neq 2$, due to interactions. The other essential
features of general filling, i.e., the energy level crossings as the
ground state symmetry changes abruptly, are qualitatively the same as
that for one atom. For many particles this is a non-trivial result
indicative of quantum phase transitions.

Fig.~\ref{fig:5} contrasts results for five particles with those for
four particles. This situation is qualitatively different as there are
no jumps in the energy derivative, or changes in the symmetry of the
ground state. Instead of exact level crossings, avoided crossings are
observed. This is reflected in the smoothed transitions in
$\Lambda_C/n_ C$. The reason is straightforward and illuminating,
since for four particles in a $4\times4$ lattice the application of
${\cal R}(2\pi/4)$ one time leaves the wave function unchanged. In
this case, the ground state symmetry must be four-fold
degenerate. Since the ground state and excited states have the same
symmetry they can mix and curve crossings do not occur. For
2-particles in a $4\times4$ lattice, the eigenvalues of ${\cal
R}(2\pi/4)$ are instead two-fold degenerate, but since the symmetry
always switches at curve crossing points, this exhibits quantum phase
transitions qualitatively like the five particle case discussed
previously.

Based on symmetry considerations, we identify transitions between
rotational ground states as second order quantum phase transitions. As
$\Omega$ increases, the state of the system changes continuously while
the symmetry of the ground state changes discontinuously.  This
defines a second order phase transition~\cite{landau1969}. Although we
have considered small quantum systems, the number of symmetry
eigenvalues $j=4$ remains true for a square lattice of arbitrary size.
Identical rotational constraints point to the same symmetry values for
the rotating 3D lattice although the exact nature of transitions will
depend on dimensionality.

We note that the appearance of a vortex lattice in the discrete system
should be observable by expansion and interference with a non-rotating
system, as in the continuous case~\cite{madison2000}.  A second
observable has been provided by our calculations: the current on the
boundary jumps discontinuously with increasing $\Omega$. The maximal
value of $\Lambda$ at eigenvalue $m=3$, indicates a current pattern
reminiscent of a tightly packed vortex lattice.

We thank John Cooper, Meret Kraemer, Erich Mueller, Brandon Peden,
Brian Seaman, and David Wood for useful discussions. We acknowledge
the support of the Department of Energy, Office of Basic Energy
Sciences via the Chemical Sciences, Geosciences and Biosciences
Division.


\end{document}